\documentclass[journal,twoside]{IEEEtran}

\usepackage{array}
\usepackage{graphicx}
\usepackage{epstopdf}
\usepackage{ifpdf}
\usepackage{subfig}
\usepackage[font=scriptsize]{caption}
\usepackage{amssymb}
\usepackage{amsmath}
\usepackage{mathrsfs}
\usepackage{longtable}
\usepackage{tabularx}
\usepackage{blindtext}
\usepackage{cite}
\usepackage{color}
\usepackage{url}
\usepackage{algorithm}
\usepackage{algorithmicx, algpseudocode}
\usepackage{balance}

\newcommand{\la}{\langle}
\newcommand{\ra}{\rangle}

\DeclareMathOperator{\loga}{\mathrm{log}}

\newcommand{\ds}{\displaystyle}
\newcommand{\st}{{\sf s.t.}}
\newcommand{\x}{{\sf x}}

\newcommand{\Psum}[1]{P_{\mathsf{#1}}}

\begin{document}

\title{Energy Efficiency in Cell-Free Massive MIMO with Zero-Forcing Precoding Design }

\author{Long~D.~Nguyen, Trung~Q.~Duong, Hien~Q.~Ngo, and Kamel Tourki
\thanks{This work was supported in part by the U.K. Royal Academy of Engineering Research Fellowship under Grant RF1415$\backslash$14$\backslash$22 and by U.K. EPSRC under Grant EP/P019374/1.}
\thanks{L. D. Nguyen, T. Q. Duong, and H. Q. Ngo are with Queen's University Belfast, Belfast, U.K. (email: \{lnguyen04, trung.q.duong, h.ngo\}@qub.ac.uk)}
\thanks{Kamel Tourki is with France Research Center, Huawei Technologies Co. (e-mail: kamel.tourki@huawei.com)}
\vspace{-0.9cm}}
\maketitle
\vspace{1cm}

\begin{abstract}
We consider the downlink of a cell-free massive multiple-input multiple-output (MIMO) network where numerous distributed access points (APs) serve a smaller number of users under time division duplex operation. An important issue in deploying cell-free networks is high power consumption, which is proportional to the number of APs. This issue has raised the question as to their suitability for green communications in terms of the total energy efficiency (bits/Joule). To tackle this, we develop a novel low-complexity power control technique with zero-forcing precoding design to maximize the energy efficiency of cell-free massive MIMO taking into account the backhaul power consumption and the imperfect channel state information. 
\end{abstract}
\vspace{-0.2cm}
\section{Introduction}

Cell-free massive multiple-input multiple-output (MIMO) has been considered as a potential technology in 5G communication, for its ability to offer uniformly good service to all users \cite{HienJ2017}. In cell-free massive MIMO, a number of single-antenna users are randomly located in a wide network area, and are coherently served by numerous low power and distributed access points (APs) \cite{HienJ2017, Interdonato2016}. Cell-free massive MIMO can be implemented with simple linear processing such as conjugate beamforming (CB) \cite{HienJ2017, Nayebi2015} and zero-forcing (ZF) \cite{Nayebi2015}. The CB technique is very simple, and has low backhaul requirements \cite{HienJ2017, Interdonato2016}. However, it suffers from high inter-user interference. By contrast, ZF processing has higher implementation complexity with higher backhaul requirement, but it can deal with the inter-user interference \cite{Nayebi2015}.

By exploiting a massive number of APs in large-scale networks, energy efficiency (EE) performance in terms of bits/Joule is a major figure-of-merit which has been neglected in most of previous works on cell-free networks \cite{HienJ2017,Interdonato2016}. It is therefore of paramount importance to design precoding techniques improving the EE performance of cell-free massive MIMO. The largest proportion of total utilized energy is exploited at APs for data transmission power, power consumption of circuit and the power consumption via backhaul network \cite{Dai2016}. Hence, it is critical to meet the ratio of green transmitted bits and total power consumption as a definition of EE performance. Furthermore, the transmitted bits need to satisfy the quality-of-service (QoS) which guarantees the minimum spectral efficiency.

In this paper, we consider a low-complexity ZF precoding design which handles the inter-user interference in cell-free massive MIMO. By our proposed process, the EE maximization problem is low-complexity and only forms as a power allocation problem. We formulate an EE maximization problem with multiple constraints where we propose simple path-following algorithms. Our proposed algorithms only need a few iterations to converge to a locally optimal solution.

\emph{Notation}: Boldface upper and lowercase letters denote matrices and vectors, respectively. The transpose complex conjugate and conjugate-transpose of matrix $\pmb{X}$ are respectively represented by $\pmb{X}^T$, $\pmb{X}^*$ and $\pmb{X}^H$. $\pmb{I}_M$ stands for the identity  matrix of size $M\times M$. $||\pmb{x}||^2 = {\la \mathbf{x},\mathbf{x}\ra}$ is the squared norm of $\mathbf{x}$. ${\sf diag}(\pmb{x})$ defines a diagonal matrix with elements of $\pmb{x}$ on its diagonal. ${\sf diag}(\pmb{X})$ denotes a vector of diagonal elements of $\pmb{X}$. $\pmb{x}_{[:,j]}$ and $\pmb{x}_{[i,:]}$ represent the $j$th column and $i$th row of matrix $\pmb{X}$. A Gaussian random vector with mean $\bar{\pmb{x}}$ and covariance $\pmb{R}_{\pmb{x}}$ is denoted by $\pmb{x}\sim \mathcal{CN}(\bar{\pmb{x}},\pmb{R}_{\pmb{x}})$.
\vspace{-0.2cm}
\section{System Model and Formulation Problem}

\subsection{System Model}
We consider a cell-free massive MIMO downlink where $K$ single-antenna users are served by $M$ randomly deployed single-antenna APs in the same time-frequency resource. A central processing unit (CPU) connects to all the APs via a backhaul network for exchanging the network information, i.e. the channel estimates, precoding vectors, and power control coefficients. 

Suppose that $g_{mk}$ is the channel between the $m$th AP and the $k$th user. As in \cite{Nayebi2015}, we adopt the following channel model
$g_{mk} =\sqrt{\beta_{mk}} h_{mk}$, where $\beta_{mk}$ represents the large-scale fading while $h_{mk} \in {\cal CN}(0,1)$ is the small-scaling fading. We assume the channel is reciprocal, i.e., the channel coefficients for uplink and downlink transmissions are the same. The channel matrix between all APs and users is denoted by $\pmb{G} \in \mathbb{C}^{M \x K}$. We further assume $M \gg K$ \cite{HienJ2017}.

The transmission from the APs to the users is done via time-division duplex protocol (TDD). Focus on the downlink performance, each coherence interval of length $\tau$ symbols is divided into two phases: uplink training and downlink payload data transmission. In the first phase, all users synchronously send pilot sequences to the APs. Then, from the received pilot signals, each AP estimates the channels to all users. In the second phase, the APs use the channel estimates to precode and beamform data to all users. 

\subsubsection{Uplink training}
Let $\tau_u$ be the length of coherence interval slot for the uplink training (in samples), and $\pmb{\varphi}_k \in C^{\tau_u\times 1}$ be the pilot sequence assigned for the $k$th user, $k=1, \ldots, K$. We assume that all pilot sequences are mutually orthonormal, i.e., $\pmb{\varphi}_k^H\pmb{\varphi}_j = 0$ for $k \neq j$ and $|| \pmb{\varphi}_k ||^2 = 1$, which requires $\tau_u \geq K$.       

The pilot signal received at the $m$th AP is 
\begin{eqnarray} \label{eq:y_rev_training}
\pmb{y}_{m} = \sqrt{\rho_r \tau_u} \sum_{k = 1}^K g_{mk} \pmb{\varphi}_k + \pmb{n}_m,
\end{eqnarray}
where $\rho_r$ is the normalized uplink power and $\pmb{n}_m \sim {\cal CN}(0, \pmb{I}_{\tau_u})$ is additive noise. The $m$th APs uses the received pilots (\ref{eq:y_rev_training}) and the minimum mean squared error (MMSE) technique to estimate the channel $g_{mk}$. Let us denote $\hat{g}_{mk}$ be the channel estimation of $g_{mk}$, then $\hat{\pmb{G}} \in \mathbb{C}^{M \x K}$ is the matrix channel estimation of $\pmb{G}$. Let $\tilde{g}_{mk} = g_{mk} - \hat{g}_{mk}$ be the channel estimation error. With the MMSE channel estimation scheme, $\hat{g}_{mk}$ and $\tilde{g}_{mk}$ are independent \cite{Nayebi2015}, and 
\[
\hat{g}_{mk} \sim {\cal CN}(0, \frac{\rho_r \tau_u \beta_{mk}^2}{1 + \rho_r \tau_u \beta_{mk}}),
\] 
\[
\tilde{g}_{mk} \sim {\cal CN}(0, \beta_{mk} - \frac{\rho_r \tau_u \beta_{mk}^2}{1 + \rho_r \tau_u \beta_{mk}}).
\]

\subsubsection{Downlink Payload Data Transmission}

The transmitted signal of $m$th AP to users is given by
\begin{eqnarray} \label{eq:x_trans}
x_m = \sqrt{\rho_f} \sum_{k = 1}^K \bar{f}_{mk} s_{k},
\end{eqnarray}
where $\rho_f$ is the downlink power of each AP. $\bar{f}_{mk}$ are the precoding coefficients, satisfying $E\{|| {x}_m ||^2\} \leq \rho_f$, and $s_k$, where $E\{|s_{k}|^2\} = 1$, is the symbol intended for the $k$th user. The received signal at the $k$th user is given by
\begin{eqnarray} \label{eq:y_downlink}
y_k = \sum_{m = 1}^M g_{mk} x_{m} + n_k,
\end{eqnarray}
where $n_k\in {\cal CN}(0,1)$.

\subsubsection{Zero-Forcing Precoding Design}
We use ZF precoding for the downlink transmission. If the APs have perfect knowledge of the channel state information (CSI), the inter-user interference can be eliminated by ZF technique.
 
With ZF processing, $\bar{f}_{mk}$ in (\ref{eq:x_trans}) can be expressed as \cite{Nayebi2015}  
\begin{eqnarray} 
\bar{f}_{mk} = \sqrt{\eta_{k}} b_{mk} , \ \ m = 1, ..., M \, ,  k = 1, ..., K,
\end{eqnarray} 
where $\eta_k, k = 1, \ldots ,K$ are the power control coefficients, and $b_{mk}$ is the $(m,k)$th element of $\pmb{B}$, where $\pmb{B} = \hat{\pmb{G}}^*(\hat{\pmb{G}}^T \hat{\pmb{G}}^*)^{-1} \in \mathbb{C}^{M \x K} $.
Let $\bar{\pmb{F}}$ be the precoding matrix whose $(m,k)$th element is $\bar{f}_{mk}$. Then, the precoding matrix $\bar{\pmb{F}}$ can be represented as
\begin{eqnarray} 
\bar{\pmb{F}} = \pmb{B} \pmb{P} ,
\end{eqnarray}
where $\pmb{P}$ is a diagonal matrix with $[\pmb{P}]_{kk} = \sqrt{\eta_{k}} , k = 1, ..., K$. Similarly to \cite{Nayebi2015}, the received signal at the $k$th user is
\begin{eqnarray} \label{eq:y_rev_final}
y_k &=& \sqrt{\rho_f} \pmb{g}_{[:,k]}^T \bar{\pmb{F}} \pmb{s} + n_k \nonumber \\ 
	&=& \sqrt{\rho_f} (\hat{\pmb{g}}_{[:,k]} + \tilde{\pmb{g}}_{[:,k]})^T \pmb{B} \pmb{P} \pmb{s} + n_k \nonumber \\
	&=& \sqrt{\rho_f} \sqrt{\eta_{k}} {s}_k + \sqrt{\rho_f} \tilde{\pmb{g}}_{[:,k]}^T \pmb{B} \pmb{P} \pmb{s} + n_k.
\end{eqnarray}
The first term of (\ref{eq:y_rev_final}) is the desired signal while the second term is the interference caused by the channel estimation error. 

Since the channel estimation is taken into account, we have to look at the spectral efficiency which includes the channel estimation overhead. Let $\pmb{\eta} = [\eta_1, \ldots, \eta_K]^T$. Then, the spectral efficiency of the $k$th user using ZF precoding is given by
\begin{eqnarray} \label{eq:rate} 
r_k(\pmb{\eta}) = (1 - \frac{\tau_u}{\tau}) \loga_2 \left(1 + \frac{\rho_f \eta_k}{ 1 + \rho_f \sum_{i=1}^K \gamma_{ki} \eta_i }\right),
\end{eqnarray}
where $\gamma_{ki} $ is the $i$th element of $\pmb{\gamma}_k$ is given by
\begin{eqnarray} \label{eq:chan_est}
 \pmb{\gamma}_k = {\sf diag}\{ E\left(\pmb{B}^H E(\tilde{\pmb{g}}_{[:,k]}^* \tilde{\pmb{g}}_{[:,k]}^T) \pmb{B}\right) \}.
\end{eqnarray}

Hence, the sum spectral efficiency is
\begin{eqnarray} \label{eq:sum_rate}
	r(\pmb{\eta}) = \sum_{k=1}^K r_k(\pmb{\eta}). 
\end{eqnarray}

To satisfy the power constraint at each AP, i.e., $E\{|| {x}_m ||^2\} \leq \rho_f$, we have 
\begin{eqnarray} \label{eq:pow_cons}
\sum_{i=1}^K \theta_{mi} \eta_i \leq 1, m = 1,..., M,
\end{eqnarray}
where $\theta_{mi}$ is the $i$th element of $\pmb{\theta}_{[m,:]}$ with $\pmb{\theta}_{[m,:]} = {\sf diag}\left\{ E\left( (\hat{\pmb{G}}^T \hat{\pmb{G}}^*)^{-1} \hat{\pmb{g}}_{[m,:]}^T \hat{\pmb{g}}_{[m,:]}^* (\hat{\pmb{G}}^T \hat{\pmb{G}}^*)^{-1} \right) \right\}$.

The total power consumption for the downlink transmission is given by \cite{Dai2016}
\begin{align} \label{eq:pow_total}
\Psum{total}  = P_{\sf cir} + \sum_{m = 1}^{M} P_m + \sum_{m = 1}^{M} P_{{\sf bh},m},
\end{align}
where $P_{\sf cir}$ denotes the static circuit power consumption, $P_m = \alpha_m \rho_f N_0 (\sum_{i=1}^K \theta_{mi} \eta_i) + P_{c,m}$ where $\alpha_m$  is the reciprocal of drain efficiency of the power amplifier at the $m$th AP, $N_0$ is the noise power, and $P_{c,m}$ is the internal power of circuit components requirement. In addition, $P_{{\sf bh},m}$ represents the power consumption of backhaul link which is used to transfer the data from the $m$th AP to the CPU. This power consumption can be modelled as 
\begin{align} \label{eq:pow_total_bh}
P_{{\sf bh},m}  = P_{0,m} + B \cdot r(\pmb{\eta}) \cdot P_{{\sf bt},m},
\end{align}  
where $P_{0,m}$ is fix power consumption for the $m$th backhaul link, $B$ is the bandwidth of system, and $P_{{\sf bt},m}$ is the traffic-dependent power in (Watt/bit/s). For convenience, we define $\bar{P}_{\sf fix} = P_{\sf cir} + \sum_{m = 1}^{M} (P_{c,m} + P_{0,m})$ as the total power consumption which is independent of $\{ \eta_{k} \}$. 
\subsection{Formulation problem}
The EE maximization problem is formulated as:
\begin{subequations} \label{eq:org_opt}
	\begin{align}
	\underset{\mathbf{\pmb{\eta}}}{\text{max}} \, & \ds \frac{B \cdotp r(\pmb{\eta})}{\Psum{total}(\pmb{\eta})}\quad   \label{eq:ref_org_opt1} \\
	\st \,\, & r_{k}(\pmb{\eta})  \geq \bar{r}_{k}, \,  k = 1,...,K, \label{eq:ref_org_opt2} \\
	&   \sum_{k=1}^K \theta_{mk} \eta_k \leq 1, m = 1,..., M,  \label{eq:ref_org_opt3} \\
	& \eta_k \geq 0 \ ,  k = 1,...,K, \label{eq:ref_org_opt4}
	\end{align}
\end{subequations}
where the constraint \eqref{eq:ref_org_opt2} represents the QoS requirement for each user. The constraint \eqref{eq:ref_org_opt4} makes sure that all the power control coefficients are positive.

The objective function in (\ref{eq:ref_org_opt1}), which is the ratio of the sum throughput and the total power consumption, represents the energy-efficient in bits/Joule. Note that the energy efficiency (\ref{eq:ref_org_opt1}) can be rewritten as
\begin{align} \label{eq:trans_org_opt}
{\sf EE}  = \frac{1}{\frac{\rho_f N_0 \sum_{m = 1}^{M} \alpha_m (\sum_{i=1}^K \theta_{mi} \eta_i) + \bar{P}_{\sf fix}}{B \cdotp r(\pmb{\eta})} + \sum_{m=1}^{M} P_{{\sf bt},m}}.
\end{align}   
Without loss of generality, maximizing EE is equivalent to minimizing the first term of the denominator of (\ref{eq:trans_org_opt}). As a result, the optimization problem (\ref{eq:org_opt}) is equivalent to maximize
\begin{align} \label{eq:trans_org_opt1}
\frac{B \cdotp r(\pmb{\eta})}{\rho_f N_0 \sum_{m = 1}^{M} \alpha_m (\sum_{i=1}^K \theta_{mi} \eta_i) + \bar{P}_{\sf fix}},  \st (\ref{eq:ref_org_opt2}), (\ref{eq:ref_org_opt3}), (\ref{eq:ref_org_opt4}).
\end{align}
In the next section, we provide the solution of problem for two cases: perfect and imperfect channel estimation at the APs.

\section{Maximizing EE with perfect channel estimation (PCE)}

Assuming perfect channel estimation, which is reasonable in the cases where the coherence interval is large (corresponding to the scenarios with low terminal mobility), we have $\hat{g}_{mk} = {g}_{mk}$ or $\tilde{g}_{mk} = 0$, and thus, the second term in (\ref{eq:y_rev_final}) is removed. 

From (\ref{eq:trans_org_opt1}), the EE maximization problem can be rewritten as
\begin{subequations} \label{eq:opt_1}
	\begin{align}
	\underset{\mathbf{\pmb{\eta}}}{\text{max}} \, & \ds \frac{B \sum_{k=1}^K (1 - \frac{\tau_u}{\tau}) \loga_2 (1 + \rho_f \eta_k)}{\rho_f N_0 \sum_{m = 1}^{M} \left(\alpha_m (\sum_{i=1}^K \theta_{mi} \eta_i) \right) + \bar{P}_{\sf fix}}   \label{eq:ref_opt1} \\
	\st \,\, & (\ref{eq:ref_org_opt2}), (\ref{eq:ref_org_opt3}), (\ref{eq:ref_org_opt4}) \label{eq:mod_opt_2}.
	\end{align}
\end{subequations} 
We can see that the objective function in (\ref{eq:ref_opt1}) is a ratio of concave and affine functions while the constraints in (\ref{eq:mod_opt_2}) are convex. Therefore, the problem (\ref{eq:opt_1}) can be solved by Dinkelbach's algorithm for fractional programming \cite{dinkelbach1967}, which find the optimal value as $\lambda > 0$ such that zero is the optimal value of the following convex program
\begin{equation} \label{eq:mod_opt_1}
\ds\max_{\pmb{\eta}, \lambda}\ B \sum_{k=1}^{K} (1 - \frac{\tau_u}{\tau}) \loga_2 (1 + \rho_f \eta_k) - \lambda \Psum{total}(\pmb{\eta})\
\mbox{s.t.}\ (\ref{eq:mod_opt_2}).
\end{equation}
The algorithm solves (\ref{eq:mod_opt_1}) using Dinkelbach's method follows \cite{LongSIP2016}.

\section{Maximizing EE with imperfect channel estimation (IPCE)}

In practice, the CSI cannot be exactly estimated. Applying MMSE estimation, the element of vector in (\ref{eq:chan_est}) with $E(\pmb{\tilde{g}}_{k}^*\pmb{\tilde{g}}_{k}^T)$ is a diagonal matrix in which the $m$th component is given by $(\beta_{mk} - \frac{\rho_r \tau_u \beta_{mk}^2}{1 + \rho_r \tau_u \beta_{mk}})$. \\

Therefore, the EE maximization problem is given by
\begin{subequations} \label{eq:opt_2}
	\begin{align}
	\underset{\mathbf{\pmb{\eta}}}{\text{max}} \, & \ds \frac{B \sum_{k=1}^K (1 - \frac{\tau_u}{\tau}) \loga_2 (1 + \rho_f \eta_k/(1 + \rho_f \sum_{i=1}^K \gamma_{ki} \eta_i) )} {\sum_{m = 1}^{M} \left(\alpha_m \rho_f N_0 (\sum_{i=1}^K \theta_{mi} \eta_i) \right) + \bar{P}_{\sf fix}}   \label{eq:ref_opt21} \\
	\st \,\, & \loga_2 (1 + \frac{\rho_f \eta_k}{1 + \rho_f \sum_{i=1}^K \gamma_{ki} \eta_i} )  \geq \tilde{r}_{k}, \, k = 1,...,K, \label{eq:ref_opt22} \\
		&   (\ref{eq:ref_org_opt3}), (\ref{eq:ref_org_opt4}),
	\end{align}
\end{subequations}
where $\tilde{r}_{k} = \frac{\tau}{\tau - \tau_u}\bar{r}_k$. Clearly, the numerator of the objective function in (\ref{eq:opt_2}) is no longer concave so the Dinkelbach's algorithm cannot be applied. We next propose an efficient procedure for solving (\ref{eq:opt_2}), which needs to solve only a few quadratic convex programs.

Following the fact that the function $f(x,t)=\frac{\ln (1+1/x)}{t}$ is convex in $x>0, t>0$ (which can be proved by examining its Hessian), the following inequality for all $x>0$, $\bar{x}>0$, $t>0$ and $\bar{t}>0$ holds \cite{Tuybook}:
\begin{eqnarray} \label{ineq4}
\ds\frac{\ln(1+1/x)}{t}\geq f(\bar{x},\bar{t})+\la \nabla f(\bar{x},\bar{t}), (x,t)-(\bar{x},\bar{t})\ra\nonumber\\
=2\ds\frac{\ln(1+1/\bar{x})}{\bar{t}}+\frac{1}{\bar{t}(\bar{x}+1)}-\ds\frac{x}{(\bar{x}+1)\bar{x}\bar{t}}-
\frac{\ln(1+1/\bar{x})}{\bar{t}^2}t.\label{ineq1}
\end{eqnarray}

By replacing $1/x$ with $x$ and $1/\bar{x}$ with $\bar{x}$, \eqref{ineq1} can be rewritten as
\begin{equation}
\ds\frac{\ln (1+x)}{t}\geq a-\frac{b}{x}-ct, \label{ineq2}
\end{equation}
where $a=2\frac{\ln(1+\bar{x})}{\bar{t}}+\frac{\bar{x}}{\bar{t}(\bar{x}+1)}>0$, $b=\frac{\bar{x}^2}{\bar{t}(\bar{x}+1)}>0$, and $c=\frac{\ln(1+\bar{x})}{\bar{t}^2}>0$.

Finally, by exploiting the fact that function $x^2/t$ is convex in $x>0$ and $t>0$, we obtain $\ds\frac{x^2}{t}\geq 2\frac{{\bar{x}}{x}}{\bar{t}}-\frac{\bar{x}^2}{\bar{t}^2}t$,  $\forall\ x>0$, $\bar{x}>0$, $t>0$, $\bar{t}>0$. 

Treating $\eta_k$ as a new variable $\eta_k^2$, problem (\ref{eq:opt_2}) is equivalent to the following quadratically constrained optimization problem:
\begin{subequations} \label{eq:ee_opt_3}
	\begin{align}
\underset{\mathbf{\pmb{\eta}}}{\text{max}} \, & \frac{B}{\ln 2} (1 - \frac{\tau_u}{\tau})  F(\pmb{\eta}) \\
\st & \quad  \rho_f \eta_{k}^2 \geq (2^{\tilde{r}_{k}}-1) (1 + \rho_f \sum_{i=1}^K \gamma_{ki} \eta_i^2), k = 1,...,K, \\
& \quad \sum_{k=1}^K \theta_{mk} \eta_{k}^2 \leq 1, m = 1,..., M,  \eta_k \geq 0 \ ,\forall k,
	\end{align}
\end{subequations}
where 
\[
F(\pmb{\eta}) \triangleq \ds \frac{\sum_{k=1}^K \ln ({1 + \rho_f \eta_k^2}/{ (1 + \rho_f \sum_{i=1}^K \gamma_{ki} \eta_i^2) }) }{ P_{\sf total} (\pmb{\eta})}.
\]


Let $\pmb{\eta}^{(n)}$ be a feasible point for the constraints in (\ref{eq:ee_opt_3}). The use of the inequality (\ref{ineq2}) for $x = x_k = \rho_f \eta_{k}^2/(1 + \rho_f \sum_{i=1}^K \gamma_{ki} \eta_i^2)$, $t= P_{\sf total} (\pmb{\eta})$, $\bar{x} = {x}_k^{(n)} = \rho_f ({\eta}_{k}^{(n)})^2/(1 + \rho_f \sum_{i=1}^K \gamma_{ki} ({\eta}_{i}^{(n)})^2$ and $\bar{t} = {t}^{(n)} = P_{\sf total} (\pmb{\eta}^{(n)})$ yields
\begin{eqnarray} \label{eq:proof}
F(\pmb{\eta})&\geq&  F^{(n)}(\pmb{\eta}),\label{fn} 
\end{eqnarray} 
where
\begin{eqnarray}
F^{(n)}(\pmb{\eta}) &\triangleq
\ds\sum_{k=1}^K\left[a^{(n)}_{k} - \ds\frac{b^{(n)}_{k}} {\rho_f \eta_{k}^2} - \ds b^{(n)}_{k} \rho_f \sum_{i=1}^K [\frac{2 \gamma_{ki}  \eta_i^{(n)}}  {\rho_f ({\eta}_{k}^{(n)})^2} \eta_i \right. & \nonumber\\
& - \ds\frac{\gamma_{ki} ({\eta}_{i}^{(n)})^2}  {(\rho_f ({\eta}_{k}^{(n)})^2)^2} \rho_f \eta_{k}^2  ]
\left. - c_{k}^{(n)}P_{\sf total} (\pmb{\eta})\right],& \label{ineq21}
\end{eqnarray}
and $0<a^{(n)}_{k}\triangleq 2\ds\frac{\ln(1+x^{(n)}_{k})}{t^{(n)}} + \frac{x^{(n)}_{k}}{t^{(n)} (x^{(n)}_{k}+1)}$,
$0<b^{(n)}_{k}\triangleq \ds\frac{(x^{(n)}_{k})^2}{t^{(n)} (x^{(n)}_{k}+1)}$, and
$0<c^{(n)}_{k}\triangleq \ds\frac{\ln(1+x^{(n)}_{k})}{(t^{(n)})^2}, k=1,\dots, K$. The proof of (\ref{eq:proof}) follows the proof in \cite{LongAccess2016}. The initial point $\pmb{\eta}^{(0)}$ can be easily determined because the constraints in (\ref{eq:ee_opt_3}) are convex.

\section{Numerical Results}

In this section, numerical results are provided to evaluate the EE performance of the considered cell-free massive MIMO system and highlight the advantage of our proposed optimization solutions. We consider an area of 1x1 km$^2$ with wrapped-around technique to avoid the boundary effects. All APs and users are distributed randomly within the area. The coefficient $\beta_{mk}$ models the large-scale fading as COST Hata model \cite{Nayebi2015}
\begin{eqnarray} \label{eq:path_loss}
\beta_{mk} = 10^{-13.6 -3.5\loga_{10}(d_{mk}) + X_{mk}/10},
\end{eqnarray}   
where $d_{mk}$ is the distance between the $m$th AP and the $k$th user in kilometers. The quantity $10^{(X_{mk}/10)}$ represents the shadowing effect with $X_{mk} \sim {\cal N} (0, \sigma_{\sf shad}^2)$. We choose $\sigma_{\sf shad} = 8$ dB, the carrier frequency $f_c = 1.9$ GHz, and bandwidth $ B = 20$ MHz. Furthermore, we choose $\tau = 200$ and $\tau_u = K$ samples. The maximum transmit power of each AP ($\rho_f$) and user ($\rho_r$) are $200$ and $100$ mW. The noise power at the receivers is $N_0 = 290 \ \x \ \kappa \ \x \ B \ \x \ NF$, where $\kappa$ and $NF$ are Boltzmann constants and noise figure at $9$ dB, respectively.
The power consumption parameters are provided similarly as in \cite{Bjornson2015, Zuo2017}. The drain efficiency of amplifier is set as $\alpha_m = 1/0.388$. The internal circuit power and the static circuit power are chosen as $P_{cm} = 0.2$ W and $P_{\sf cir} = 9$ W. For backhaul power consumption, we choose the fixed power backhaul link and the traffic dependent backhaul power as $P_{0,m} = 0.2$ W and $P_{{\sf bt},m} = 0.25$ W/(Gbits/s), respectively.
\vspace{-0.4cm}    
\begin{figure}[H]
	\centering
	\centerline{\includegraphics[width=0.35\textwidth]{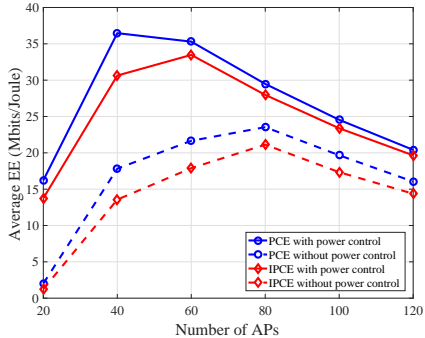}}
	\caption{The average EE performance versus the number of APs (M). $K = 16$.}
	\label{fig:1_APs}
\end{figure}
\vspace{-0.2cm}
For comparison, we also provide the case without power control, i.e., when each AP uses its power transmission such that all power coefficients ($\eta_k$) equal to $1/(\underset{\{m\}}{\sf max}(\sum_{k = 1}^K \theta_{mk}))$. The QoS constraint is set to be equal to the spectral efficiency in this case.

As can be clearly seen from Fig. 1, our proposed scheme outperforms the equal power allocation in terms of EE performance for both PCE and IPCE cases. For the case without power control, the optimal performance can be achieved at $M=80$ whereas with our proposed algorithms, the optimal performance can be achieved at $M=40$ and $M=60$ for PCE and IPCE, respectively. It is interesting to see that the use of more APs beyond these optimal points does not improve the EE performance as the power consumption level also increase with the number of APs.

In Fig. 2, we demonstrate the network EE performance between different schemes versus the transmit power at each AP from $0.2$ to $2.2$ W. The EE performance increases noticeably with $\rho_f$ when $\rho_f$ is small ($< 1$ W), and saturates when the transmit power is greater than $1$ W. The reason comes from the fact that when the transmit power is high, we are in interference-limited regimes and domination of transmission power, and hence, we cannot improve the system performance by simply increasing the transmitted power.
\vspace{-0.4cm}
\begin{figure}[H]
	\centering
	\centerline{\includegraphics[width=0.35\textwidth]{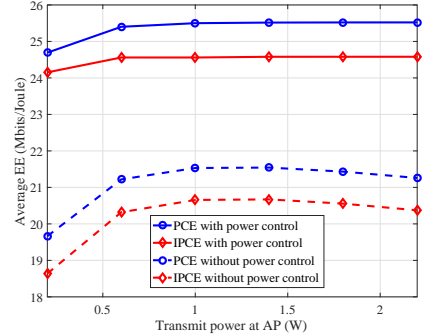}}
	\caption{The average EE performance versus the transmit power at AP ($\rho_f$). $M = 100$, $K = 16$, $\bar{r}_k = r = 1$ bits/s/Hz.}
	\label{fig:2_Pf}
\end{figure}
\vspace{-0.8cm}
\section{Conclusion}
We have proposed new algorithms with low complexity for maximizing the energy efficiency of zero-forcing precoding in the downlink transmission of cell-free massive MIMO while satisfying per-user QoS constraints and per-AP transmit power constraint. In addition, for the case of imperfect CSI, the pathfollowing algorithm has introduced which are more tractable and applicable than the Dinkelbach’s approach (i.e., only suitable for perfect CSI). The numerical results have demonstrated the effectiveness of the power control algorithms, compared with no power control.


\bibliographystyle{IEEEtran}
\bibliography{References}

\end{document}